# WESSON'S INDUCED MATTER THEORY

# WITH A WEYLIAN BULK.

Mark Israelit[1]

*The foundations of Wesson's Induced Matter Theory are analyzed. It is shown that the empty, - without matter, - 5-dimensional bulk must be regarded as a Weylian space rather than as a Riemannian one. Revising the geometry of the bulk, we have assumed that a Weylian connection vector and a gauge function exist in addition to the metric tensor. The framework of a Weyl-Dirac version of Wesson's theory is elaborated and discussed. In the 4-dimensional hypersurface (brane), one obtains equations describing both fields, the gravitational and the electromagnetic. The result is a geometrically based unified theory of gravitation and electromagnetism with mass and current induced by the bulk. In special cases on obtains on the brane the equations of Einstein-Maxwell, or these of the original Induced Matter Theory.*

---

[1] Department of Physics and Mathematics, University of Haifa-Oranim, Tivon 36006 Israel.
e-mail: <israelit@macam.ac.il>

# 1. INTRODUCTION

In the present paper, we show that in Wesson's Space-Time-Matter theory [1] (STM) the 5-dimensional manifold (bulk) is a Weyl space rather than a Riemannian one. From the cognitive point of view, the main achievement of the STM, known also as the Induced Matter Theory (IMT), is the successful explanation of the geometric origin of matter. Therefore, we consider first some points concerning the geometry ⇔ matter interdependence.

Matter and field are basic concepts of classical field theories. They play a fundamental role in the general relativity theory [2], where the Einstein tensor $G_\mu^\nu$ is expressed in terms of the geometry of space-time, and the matter is represented by its momentum-energy density tensor $T_\mu^\nu$. These two intrinsic concepts are connected by the Einstein field equation

$$G_\mu^\nu = -8\pi\, T_\mu^\nu . \tag{1}$$

According to Eq. (1), a given distribution of matter (-sources) determines the geometric properties of space-time. One can regard this as the creation of space-time geometry by matter. Now, one can read Eq. (1) in the opposite direction, and expect for the creation of matter by geometry. The existence of this reverse process would fulfill one of Einstein's dreams. Sixty-six years ago Einstein and Infeld [3] wrote: "Could we not reject the concept of matter and build a pure field physics? … We could regard matter as the region where the field is extremely strong."

However, what has brought matter into being? There are many interesting works dealing with this problem. In the fifties Wheeler [4] introduced the concept of electromagnetic geons, massive entities created by spatially confined fields. Later

Wheeler and Power [5] introduced the thermal geon. Geons were included by Misner and Wheeler also in their Geometrodynamics [6], which describes gravitation and electromagnetism in terms of geometry.

During the last 50 years the origin of mass was discussed in many interesting works. In 1974 G. 't Hooft [7] showed that in gauge theories in which the electromagnetic group is a subgroup of a larger group, like SU(2) or SU(3), massive magnetic monopoles can be created as regular solutions of the field equations. Later, Gross and Perry [8] found in the framework of a 5-dimensional Kaluza-Klein theory regular static and stable soliton solutions, which correspond, upon quantization, to particles; these solitons include massive magnetic monopoles.

There must be given attention to classical gauge theories, where the electromagnetic or the Yang-Mills field arises from connections on U(1) or SU(2) – principle bundles over the space-time manifold {M}, while the gravitational field arises on the GL(4) principal frame bundle over {M}. One can adopt the standpoint that all non-gravitational fields should somehow uniquely determine the gravitational field and its mass [9]. Among other interesting theories that appeared in the middle of the eighties, there is a successful elegant Yang-Mills field theory of gravitation presented by H. Dehnen et al. [10]. This theory is based on a unitary phase gauge invariance of the Lagrangian, where the gauge transformations are those of SU(2) x U(1) symmetry of the two spinors. In the classical limit this microscopic theory results in Einstein's theory of gravitation.

Some years ago in an interesting work C. S. Bohum and F. I. Cooperstock [11] issued from the Lagrangian of quantum electrodynamics. For the case **A**=0, $\varphi \neq 0$ they obtained the stationary Dirac equation and the Poisson equation. These equations lead to a Dirac-Maxwell soliton with the mass and charge of the electron.

Recently classical models of elementary particles were presented by O. Zaslavskii [12]. These are built up by gluing the Reissner - Nordstrǿm metric (or the Kerr – Newman black hole) to a static (or rotating) Bertotti- Robinson core.

Now, according to classical general relativity the most attractive scenario would be creation of massive matter by geometry. However, as it was shown in the nineties by Cooperstock et al. [13, 14] no gravitational geon can exist in general relativity based on Riemannian geometry. Thus, in order to get matter creation by geometry we must turn to a more general framework than the four dimensional (4D) Riemannian one. There exist various extensions of Einstein's framework possessing massive matter stemming from pure geometry. Some years ago, the present writer proposed an integrable Weyl-Dirac theory [15]. This has to be regarded as a version of the Weyl-Dirac (W-D) theory, which is an elegant 4D unification of Einstein's GTR and electromagnetism. The enlarged W-D theory is based on Weyl's geometry [16, 17] modified by Dirac's action principle [18], and by Rosen's approach [19], the latter allowing to regard the Weyl vector field as creator of massive particles. It is worth noting that a recently proposed Weyl-Dirac Torsional Massive Electrodynamics [20] possesses massive photons and magnetic monopoles.

In the Integrable Weyl-Dirac theory [15, 21], a spatially confined, spherically symmetric formation made of pure geometric quantities is a massive entity [22]. At the very beginning the matter universe was created by an embryonic egg made of pure geometry. After the matter universe was born, the geometry stimulates cosmic matter production during the expansion phase and form dark matter and quintessence in the accelerating universe [21, 23]. Ten years ago Novello [24] proposed a different approach of Weyl's geometry the Weyl Integrable Space-Time (WIST) that leads to interesting results in cosmology.

Another extension of Einstein's theory is based on the idea that our 4D space-time is a surface embedded in a 4+$m$ ($m>0$) dimensional manifold. Proceeding from this idea Kaluza [25] proposed a unification of electromagnetism and gravitation in the frame of a 4D hypersurface embedded in a 5-dimensional (5D) manifold. Suggesting that the fifth dimension has a circular topology Klein [26] imposed the cylindricity condition and completed the Kaluza theory. In the Kaluza-Klein theory, the fifth coordinate $x^4$ plays a purely formal role and the components of the 5D metric tensor do not depend on $x^4$. In 1938 Einstein and Bergmann [27, 28] presented a generalization of the Kaluza-Klein theory. In this work the condition of cylindricity (that is equivalent to the existence of a 5D Killing vector) is replaced by the assumption that with regard to the fifth coordinate the space is periodically closed. In the Einstein - Bergmann version the fifth dimension has a physical meaning.

The Kaluza-Klein idea of extra dimensions, where ordinary matter is confined within a lower dimensional surface, has received an enormous amount of attention during the last decades. There must be noted the early works of Joseph [29], Akama [30], Rubakov and Shaposhnikov [31], Visser [32]. The basic works of Randall and Sundrum [33], as well the works of Arkani-Hamed et al [34], who suggested that ordinary matter would be confined to our 4D universe, while gravity would "live" in the extended 4+$m$ dimensional manifold, played a key role in the further development of Kaluza-Klein theories. A list of relevant papers is given in Rubakov's recent review [35]. .

On a revised Kaluza-Klein approach is based Wesson's theory [1], in which the physical matter of the 4D space-time is created by the geometry of a 5D bulk. Basic concepts and approaches of this remarkable theory as well applications to cosmology were developed during the last ten years in collaboration by Liko, Lim, Liu,

Overduin, Ponce de Leon, Seahra and Wesson [1, 36-50]. Between important achievements of the Induced Matter Theory is the proof of the geometric origin of matter, successful cosmological models with a variable cosmological constant, dark matter, accelerated universes. Wesson's framework includes also an induced unified theory of gravitation and electromagnetism.

Now, comparing Wesson's IMT with the Weyl-Dirac theory one recognizes similar results. Both theories allow getting matter from geometry, as well obtaining singularity-free cosmological models. In both theories, Wesson's IMT and the Integrable W-D theory; dark matter and quintessence follow from geometry. Both frameworks provide a unified, geometrically based description of gravity and electromagnetism. Finally, both, the Weyl-Dirac theory and the Kaluza-Klein one originate from attempts of building up unified theories of classical fields. Below it is shown that the 5-dimensional empty bulk must be described by Weyl's geometry.

In the present work, the following conventions are valid. Uppercase Latin indices run from 0 to 4; lowercase Greek indices run from 0 to 3. Partial differentiation is denoted by a comma (,), Riemannian covariant 4D differentiation by a semicolon (;), and Riemannian covariant 5D differentiation by a colon (:). Weylian derivative in a 5D space is written as $\nabla_A$, and in 4 dimensions one writes $\nabla_\mu$.

## 2. EMBEDDING A 4D SPACE-TIME IN A 5D MANIFOLD. THE FORMALISM

This section contains a concise description of the general embedding formalism. The notations as well as the geometric construction given below accord to these given in recent works of Sanjeev S. Seahra and Paul Wesson (cf. [47, 48]).

One considers a 5-dimensional manifold { $M$ } (the "bulk") with a symmetric metric $g_{AB} = g_{BA}$, ($A$, $B$ = 0, 1, 2, 3, 4) having the signature $\text{sig}(g_{AB}) = (-,+,+,+,\varepsilon)$ with $\varepsilon = \pm 1$. The manifold is mapped by coordinates { $x^A$ } and described by the line-element

$$dS^2 = g_{AB} dx^A dx^B . \qquad (2)$$

One can introduce a scalar function $l = l(x^A)$ that defines the foliation of {$M$} with 4-dimensional hypersurfaces $\Sigma_l$ at a chosen $l$ = const, as well the vector $n^A$ normal to $\Sigma_l$. If there is only one timelike direction in {$M$}, it will be assumed that $n^A$ is spacelike. If {$M$} possesses two timelike directions ($\varepsilon = -1$), $n^A$ is a timelike vector. Thus, in any case $\Sigma_l$ (the "brane") contains three spacelike directions and a timelike one. The brane (our 4-dimensional space-time) is mapped by coordinates { $y^\mu$ }, ($\mu = 0,1,2,3.$) and has the metric $h_{\mu\nu} = h_{\nu\mu}$ with $\text{sig}(h_{\mu\nu}) = (-,+,+,+)$. The line-element on the brane is

$$ds^2 = h_{\mu\nu} dy^\mu dy^\nu . \qquad (3)$$

It is supposed that the relations $y^\nu = y^\nu(x^A)$ and $l = l(x^A)$, as well as the reciprocal one $x^A = x^A(y^\nu, l)$ are mathematically well-behaved functions. Thus, the 5D bulk may be mapped either by $\{x^A\}$ or by $\{y^\nu, l\}$. The normal vector to $\Sigma_l$ is given by

$$n_A = \varepsilon \Phi \partial_A l; \tag{4}$$

with $\Phi$ being the lapse function.

A 5D quantity (vector, tensor) in the bulk has 4D counterparts located on the hypersurfaces. These counterparts may be formed by means of the following system of basis vectors, which are orthogonal to $n_A$

$$e_\nu^A = \frac{\partial x^A}{\partial y^\nu} \quad \text{with} \quad n_A e_\nu^A = 0. \tag{5}$$

Thus, the brane $\Sigma_l$ is stretched on four five-dimensional basis vectors $e_\nu^A$. Together with the main basis $\{e_\nu^A; n_A\}$ one can consider its associated one $\{e_A^\nu; n^A\}$, which also satisfies an orthogonality condition $e_A^\nu n^A = 0$. The main basis and its associated are connected by the following relations:

$$e_\nu^A e_A^\mu = \delta_\nu^\mu \;;\; e_\sigma^A e_B^\sigma = \delta_B^A - \varepsilon n^A n_B \;;\; n^A n_A = \varepsilon \;. \tag{6}$$

Let us consider a 5D vector $V_A$ in the bulk $\{M\}$. Its 4D counterpart on the brane $\Sigma_l$ is given by

$$V_\mu = e_\mu^A V_A \;;\; V^\nu = e_B^\nu V^B. \tag{7}$$

On the other hand the 5D vector may be written as

$$V_A = e_A^\mu V_\mu + \varepsilon(V_S n^S) n_A; \quad V^A = e_\mu^A V^\mu + \varepsilon(V^S n_S) n^A. \tag{8}$$

Actually, (8) is a decomposition of $V_A$ into a 4-vector $V_\mu$ and a part normal to $\Sigma_l$.

Further, the 5D metric tensor, $g_{AB}$; $g^{AB}$, and the 4D one, $h_{\mu\nu}$; $h^{\mu\nu}$, are related by

$$h_{\mu\nu} = e_\mu^A e_\nu^B g_{AB} \;;\; h^{\mu\nu} = e_A^\mu e_B^\nu g^{AB} \;;\; \text{with} \quad h_{\mu\nu} h^{\lambda\nu} = \delta_\mu^\lambda, \tag{9}$$

and

$$g_{AB} = e_A^\mu e_B^\nu h_{\mu\nu} + \varepsilon n_A n_B \;;\; g^{AB} = e_\mu^A e_\nu^B h^{\mu\nu} + \varepsilon n^A n^B \;;\; \text{with} \quad g_{AB} g^{CB} = \delta_A^C. \tag{10}$$

Details may be found in [1, 47, 48].

## 3. PARALLEL DISPLACEMENT IN WEYL'S GEOMETRY

In the next section parallel displacement on a brane in the IMT will be considered. Here we recall the cardinal idea of Weyl [16, 17]. He issued from the dominance of light rays for physical experiments. Accordingly, the light cone is the principal phenomenon describing the 4D space-time. This idea brought Weyl to regard the isotropic interval $ds^2 = 0$ as invariant rather than an arbitrary line-element $ds^2 = h_{\alpha\beta} dy^\alpha dy^\beta$ between two space-time events. Thus, in the 4D Weyl geometry, the metric interval between two events as well the length of a given vector is no more constant, it depends on an arbitrary multiplier, the gauge function. In order to describe the geometry based on the invariance of light-cones Weyl introduced in addition to the metric tensor $h_{\alpha\beta} = h_{\beta\alpha}$ a length connection vector $w_\nu$. However, in Weyl's theory there was a difficulty in obtaining satisfactory equations for gravitation. 50 years later Dirac [18] revived the Weyl theory. Modifying the variational principle, he introduced a Lagrangian multiplier, involving the gauge function $\Omega(y^\lambda)$ into the action. Dirac's modification enabled to derive satisfactory equations for gravitation and electromagnetism.

In Weyl's geometry the connection is given by

$$\Gamma^\lambda_{\mu\nu} = \left\{ {}^{\lambda}_{\mu\nu} \right\} + h_{\mu\nu} w^\lambda - \delta^\lambda_\mu w_\nu - \delta^\lambda_\nu w_\mu, \qquad (11)$$

with $\left\{ {}^{\lambda}_{\mu\nu} \right\}$ being the Christoffel symbol. Let us consider a vector undergoing an infinitesimal parallel displacement $dy^\nu$, so that its component $V^\mu$ changes by

$$\underset{W}{d} V^{\mu} = -V^{\sigma}\Gamma^{\mu}_{\sigma\nu}dy^{\nu} \ . \tag{12}$$

From (11) and (12) one obtains the following change of the length $V = \sqrt{V_{\lambda}V_{\sigma}h^{\lambda\sigma}}$:

$$\underset{W}{d} V = Vw_{\nu}dy^{\nu} . \tag{13}$$

If the vector is transported by parallel displacement around an infinitesimal closed parallelogram the total change of its length according to (11) and (13) is

$$\underset{W}{\Delta} V = V W_{\mu\nu} dy^{\mu} \delta y^{\nu} \ , \tag{14}$$

where

$$W_{\mu\nu} = w_{\mu,\nu} - w_{\nu,\mu} \tag{15}$$

is the Weylian length curvature tensor, which in the Weyl-Dirac theory is identified as the electromagnetic field tensor. Thus, in this geometry $\underset{W}{\Delta} V \neq 0$ unless $W_{\mu\nu} = 0$. If a vector having the length $V_{\text{initial}}$ has been transported round a closed loop, and arrived at the starting point, the new length according to (14) will be

$$V_{\text{new}} = V_{\text{initial}} + \int_{S} V W_{\mu\nu} dS^{\mu\nu} \ , \tag{16}$$

with $S$ being the area confined by the loop, and $dS^{\mu\nu} = dy^{\mu}\delta y^{\nu}$ an element of this area. As the loop can be chosen arbitrarily, one has an arbitrary standard of length (or gauge) at each point, and one can consider local gauge transformations, called Weyl gauge transformations. Under a Weyl gauge transformation (WGT) the length of a vector changes as

$$V \to \tilde{V} = e^{\lambda}V \ , \tag{17}$$

where $\lambda(y^{\nu})$ is an arbitrary function of the coordinates; the metric tensor changes as

$$h_{\mu\nu} \to \tilde{h}_{\mu\nu} = e^{2\lambda}h_{\mu\nu} \ ; \quad h^{\mu\nu} \to \tilde{h}^{\mu\nu} = e^{-2\lambda}h^{\mu\nu} \ , \tag{18}$$

and the Weyl connection vector changes according to

$$w_\nu \to \tilde{w}_\nu + \frac{\partial \lambda}{\partial y^\nu} \ . \tag{19}$$

The gauge transformations may be represented by an arbitrary positive function, the Dirac gauge function $\Omega(y^\mu)$ that changes as

$$\Omega \to \tilde{\Omega} = e^{-\lambda}\Omega \ , \tag{20}$$

so that $\ln\left(\tilde{\Omega}/\Omega\right) = -\lambda$ and there exists a mutual correspondence between the two scalar functions, $\Omega(y^\mu)$ and $\lambda(y^\nu)$. It must be pointed out that if $w_\nu = 0$, the Weyl space turns into the Riemannian (cf. (11)). If $w_\nu$ is a gradient vector the Weylian length curvature tensor $W_{\mu\nu}$ vanishes (cf. (15)) and according to (16) one obtains an integrable space. More on the Weyl- Dirac theory may be found in [16–19, 21].

.

## 4. PARALLEL DISPLACEMENT IN THE STM THEORY

A vector $V_M$ in the 5D bulk has a 4-dim. counterpart in $\Sigma_l$, the vector $V_\mu$. These two vectors are related by equations (7) and (8). Let us consider an infinitesimal parallel displacement of $V_M$ in the bulk [2]

$$dV_M = \begin{Bmatrix} S \\ M B \end{Bmatrix} V_S \, dx^B \ . \tag{21}$$

The change of the length $(V_S V^S)^{\frac{1}{2}}$ of this vector obviously vanishes. Now, according to (7), the induced parallel displacement of $V_\mu$ is

$$dV_\mu = d(e_\mu^A V_A) = e_\mu^A dV_A + V_A de_\mu^A = e_\mu^A V_S \begin{Bmatrix} S \\ A B \end{Bmatrix} dx^B + V_A e_{\mu,B}^A dx^B \ . \tag{22}$$

---

[2] We use the curly brackets $\{\ \}$ for both, 5D Christoffel symbols and 4D ones. They may be distinguished by the indices: the 5D have uppercase Latin indices, the 4D have lowercase Greek indices

As the 5D bulk may be mapped either by $\{x^A\}$ or by $\{y^\nu, l\}$ one can write (cf. [48])

$$dx^B = e_\nu^B \, dy^\nu + l^B dl, \quad \text{with } l^B = \frac{\partial x^B}{\partial l} . \tag{23}$$

Further, the 5D vector $l^B$ may be written as

$$l^B = e_\nu^B N^\nu + \Phi n^B . \tag{24}$$

In the decomposition (24) the shift vector $N^\nu$ is parallel to $\Sigma_l$, $n^B$ stands for the normal to $\Sigma_l$, and $\Phi = \varepsilon(l^S n_S)$ is the lapse function. According to (23) and (24)

$$dx^B = e_\nu^B dy^\nu + l^B dl = e_\nu^B dy^\nu + e_\nu^B N^\nu dl + \Phi \, n^B dl . \tag{25}$$

Inserting decomposition (25) into the expression for the parallel displacement (22) we obtain

$$dV_\mu = \left(e_\mu^A V_S \left\{{}^S_{A\,B}\right\} + V_A e_{\mu,B}^A\right)\left(e_\nu^B dy^\nu + e_\nu^B N^\nu dl + \Phi n^B dl\right). \tag{26}$$

One can consider the displacement (26) as consisting of three parts:

$$dV_\mu = \underset{\Sigma_l}{d}V_\mu + \underset{l}{d}V_\mu + \underset{\perp}{d}V_\mu , \tag{27}$$

With $\underset{\Sigma_l}{d}V_\mu = \left(e_\mu^A V_S \left\{{}^S_{A\,B}\right\} + V_A e_{\mu,B}^A\right) e_\nu^B dy^\nu$; $\underset{l}{d}V_\mu = \left(e_\mu^A V_S \left\{{}^S_{A\,B}\right\} + V_A e_{\mu,B}^A\right) e_\nu^B N^\nu dl$; and

$\underset{\perp}{d}V_\mu = \left(e_\mu^A V_S \left\{{}^S_{A\,B}\right\} + V_A e_{\mu,B}^A\right) \Phi n^B dl$. The displacement $\underset{\Sigma_l}{d}V_\mu$ is located in $\Sigma_l$. Further,

$\underset{l}{d}V_\mu$ is the displacement along $N^\nu$, and the third term, $\underset{\perp}{d}V_\mu$, is normal to the 4D

brane $\Sigma_l$. For our purpose, only the term $\underset{\Sigma_l}{d}V_\mu$ is relevant and a straightforward

calculation yields

$$\underset{\Sigma_l}{d}V_\mu = V_\sigma \left\{{}^\sigma_{\mu\nu}\right\} dy^\nu + \frac{\varepsilon}{2}\left(V^S n_S\right)\left[\left(h_{\sigma\nu} e_{L,\mu}^\sigma + h_{\sigma\mu} e_{L,\nu}^\sigma\right) - \left(e_\mu^A h_{\sigma\nu} + e_\nu^A h_{\sigma\mu}\right) e_{A,L}^\sigma - h_{\mu\nu,L}\right] n^L dy^\nu .$$

$$\tag{28}$$

This may be rewritten as

$$\underset{\Sigma_l}{d}V_\mu = \left[V_\sigma \begin{Bmatrix} \sigma \\ \mu\nu \end{Bmatrix} - \varepsilon(V^S n_S)C_{\mu\nu}\right]dy^\nu, \tag{29}$$

with

$$C_{\mu\nu} = e_\mu^A e_\nu^B n_{B:A} \equiv e_\mu^A e_\nu^B \left(\frac{\partial n_B}{\partial x^A} - n_S \begin{Bmatrix} S \\ BA \end{Bmatrix}\right) \tag{30}$$

standing for the extrinsic curvature of the hypersurface $\Sigma_l$.

Now let us consider the square of the length $V^2 \equiv h^{\mu\lambda}V_\mu V_\lambda$. Its change under a parallel displacement is

$$\underset{\Sigma_l}{d}(h^{\mu\lambda}V_\mu V_\lambda) = h^{\mu\lambda}V_\mu dV_\lambda + h^{\mu\lambda}V_\lambda dV_\mu + V_\mu V_\lambda h^{\mu\lambda}_{,\nu} dy^\nu. \tag{31}$$

Making use of $h^{\mu\lambda}_{,\nu} \equiv -\begin{Bmatrix} \mu \\ \sigma\nu \end{Bmatrix} h^{\sigma\lambda} - \begin{Bmatrix} \lambda \\ \sigma\nu \end{Bmatrix} h^{\mu\sigma}$, we obtain from (31) the change of the squared length of the 4-vector:

$$\underset{\Sigma_l}{d}(V^\lambda V_\lambda) = -2\varepsilon(V_S n^S)V^\sigma C_{\sigma\nu}\, dy^\nu. \tag{32}$$

For a moment, let us go back to the 4D Weyl-Dirac geometry. From (13) we have for the change of the squared length of a vector

$$\underset{W}{d}V^2 = 2V^2 w_\nu dy^\nu. \tag{33}$$

Although relations (32) and (33) differ one from another, both describe non-integrable 4D spaces. Thus, in the general case, the 4-brane possesses a non-integrable geometry, and only when the original 5D vectors are situated entirely in $\Sigma_l$ (i.e. $V_S n^S = 0$), or when the external curvature vanishes ($C_{\mu\nu} = 0$) one has a Riemannian brane.

## 5. THE WEYL–DIRAC FORMALISM ON THE BULK

In light of the non-integrable induced geometry on 4D branes, it is justified to reconsider the geometry of the bulk. In Wesson's 5D induced matter theory, one regards the bulk as pure geometry without any additional fields. The geometry is described by the metric tensor $g_{AB}$. Thus, the principal phenomenon, which carries information is a metric perturbation propagating in the form of a gravitational wave. In order to avoid misinterpretations one must assume that all gravitational waves have the same speed. Therefore, the isotropic interval $dS^2 = 0$ has to be invariant, whereas an arbitrary line element $dS^2 = g_{AB} dx^A dx^B$ may vary. The situation resembles the 4D Weyl geometry, where the light cone is the principal phenomenon describing the space-time and hence the lightlike interval $ds^2 = 0$ is invariant rather than an arbitrary line-element $ds^2 = h_{\alpha\beta} dy^\alpha dy^\beta$ between two space-time events (cf. Sec. 3).

We will adopt the ideas of Weyl and Dirac assuming in every point of the bulk the existence of a metric tensor $g_{AB}(x^D) = g_{BA}(x^D)$ and a vector $w^A(x^D)$. The infinitesimal geometry will be described by a 5D Weylian connection

$$\Gamma_{AB}^D = \{_{AB}^D\} + \Delta_{AB}^D = \{_{AB}^D\} + g_{AB} w^D - \delta_A^D w_B - \delta_B^D w_A , \qquad (34)$$

with $\{_{AB}^D\}$ being the 5D Christoffel symbol. Consider an infinitesimal parallel displacement of a given vector $V^A$. According to (34), one writes for the change of the component

$$dV^A = -V^D \Gamma_{DB}^A dx^B , \qquad (35)$$

so that the vector's length $V = (V_S V^S)^{1/2}$ will be changed by

$$dV = V w_B dx^B . \qquad (36)$$

In the Weyl framework, if a vector $V^A$ is transported by parallel displacement round an infinitesimal closed parallelogram one finds for the total change of the components

$$\Delta V^A = V^S K^A_{SBC} dx^B \delta x^C ,  \tag{37}$$

with $K^A_{SBC}$ being the curvature tensor formed from $\Gamma^A_{BS}$:

$$K^A_{SBC} = -\Gamma^A_{SB,C} + \Gamma^A_{SC,B} - \Gamma^L_{SB}\Gamma^A_{LC} + \Gamma^L_{SC}\Gamma^A_{LB} . \tag{38}$$

From (36) one obtains the change in the length after a displacement round an infinitesimal parallelogram (cf. (14))

$$\Delta V = V W_{BC} dx^B \delta x^C , \tag{39}$$

with $W_{AB} = w_{A,B} - w_{B,A}$ being the 5-dimensional Weylian length curvature tensor.

Just as in the 4D Weyl geometry (cf. Sec. 3), so also in this 5D case one is led by equation (39) to an arbitrary standard of length. Thus, one introduces 5D Weyl gauge transformations (WGT) with a scalar function $\lambda(x^A)$ and the Dirac gauge function $\Omega(x^A)$. Under WGT one has $g_{AB} \Rightarrow \tilde{g}_{AB} = e^{2\lambda} g_{AB}$, $g^{AB} \Rightarrow \tilde{g}^{AB} = e^{-2\lambda} g^{AB}$, $\Omega \Rightarrow \tilde{\Omega} = e^{-\lambda}\Omega$ and $w_A \Rightarrow \tilde{w}_A = w_A + \lambda_{,A}$. In the 5D Weylian bulk one has two different coordinate covariant derivatives, the Weylian (cf. Appendix B, below)

$$\nabla_B V^A \equiv V^A_{,B} + V^S \Gamma^A_{SB}, \tag{40}$$

and the 5D Riemannian one

$$V^A_{:B} \equiv V^A_{,B} + V^S \left\{ {A \atop SB} \right\} . \tag{41}$$

Making use of (34) and (41) we can rewrite the length curvature tensor as

$$W_{AB} = w_{A,B} - w_{B,A} = w_{A:B} - w_{B:A} , \tag{42}$$

and the curvature tensor given by equation (38) as

$$K^A_{SBC} = R^A_{SBC} + \delta^A_S(w_{B:C} - w_{C:B}) + g_{SC}w^A_{:B} - g_{SB}w^A_{:C} + \delta^A_B w_{S:C} - \delta^A_C w_{S:B} + \\ (\delta^A_C g_{BS} - \delta^A_B g_{CS})w_L w^L + (g_{CS}w_B - g_{BS}w_C)w^A + (\delta^A_B w_C - \delta^A_C w_B)w_S . \tag{43}$$

In Eq. (43) $R^A_{SBC}$ stands for the 5D Riemannian curvature tensor formed from $\left\{ {}^{A}_{BC} \right\}$.

Contracting (43) we obtain

$$K_{MN} \equiv K^S_{MNS} = R_{MN} + w_{N:M} - 4w_{M:N} - g_{MN} w^S_{:S} + 3g_{MN} w_S w^S - 3w_M w_N \ . \tag{44}$$

From (44) one can derive

$$K_{MN} - K_{NM} = -5(w_{M:N} - w_{N:M}) = -5W_{MN} \ , \tag{45}$$

and also obtain the 5D Weylian curvature scalar

$$K^S_S = g^{MN} K_{MN} = \hat{R} - 8w^S_{:S} + 12 w_S w^S \ , \tag{46}$$

with $\hat{R} \equiv g^{AB} R^L_{ABL}$ being the 5-dimensional Riemannian curvature scalar.

Following Dirac [18] we shall derive the field equations in the 5D bulk from a variational principle $\delta \int \underset{geom}{L} \sqrt{-g}\, d^5 x = 0$, with $\underset{geom}{L} \sqrt{-g}$ being an in-invariant, i.e. invariant under both, coordinate transformations (CT) and Weyl gauge transformations (WGT).

One can form the Lagrangian $\underset{geom}{L}$ from the following suitable, geometrically based terms. 1) $\Omega W^{AB} W_{AB}$ (cf. (42)); 2) $-\Omega^3 K^S_S$ (cf. (46)); 3) regarding the Dirac gauge function $\Omega(x^A)$, in addition to $g_{AB}$ and $w_D$, as a dynamical variable one adds a term containing its gauge covariant derivatives $k \Omega g^{AB} (\Omega_{,A} + \Omega w_A)(\Omega_{,B} + \Omega w_B)$, with $k$ being an arbitrary constant; 4) finally, one adds the cosmological term $\Omega^5 \Lambda$. The $\Omega^n$ multipliers in 1) – 4) assure WGT invariance of the terms.

It is convenient to denote $\Omega_{,A} \equiv \Omega_A$; $g^{AB} \Omega_{,B} \equiv \Omega^A$. Then, making use of the above mentioned terms, 1) – 4) and discarding perfect differentials we obtain the action integral

$$I = \int \left[ \Omega W_{AB} W^{AB} - \Omega^3 \hat{R} + (k-12)\Omega^2 w^S (\Omega w_S + 2\Omega_S) + k\Omega\Omega_S \Omega^S + \Omega^5 \Lambda \right] \sqrt{-g}\, d^5 x .$$

(47)

Considering the variation of (47) with respect to $g_{AB}$, $w_D$ and $\Omega$, one obtains the field equations. It turns out that in the $\delta w_A$-equation will appear a Proca-like term if $k - 12 \neq 0$. As from the quantum mechanical standpoint, such a term may be interpreted as representing massive particles, we will assume $k = 12$. With this choice, we have the following action:

$$I = \int \left[ \Omega W_{AB} W^{AB} - \Omega^3 \hat{R} + 12\Omega\Omega_S \Omega^S + \Omega^5 \Lambda \right] \sqrt{-g}\, d^5 x . \tag{48}$$

Now, making use of (A-1) – (A-4) in Appendix A, we obtain the $\delta g^{AB}$-equation

$$R_{AB} - \frac{1}{2} g_{AB} \hat{R} = \frac{2}{\Omega^2} \left( W_{AS} W_B{}^{\cdot S} - \frac{1}{4} g_{AB} W_{LS} W^{LS} \right) + \frac{6}{\Omega^2} \Omega_A \Omega_B - \frac{3}{\Omega} \left( \Omega_{A:B} - g_{AB} \Omega^S{}_{:S} \right) - \frac{1}{2} g_{AB} \Omega^2 \Lambda , \tag{49}$$

the $\delta\Omega$ - equation

$$W_{AB} W^{AB} - 3\Omega^2 \hat{R} - 12\left( \Omega_S \Omega^S + 2\Omega\Omega^S{}_{:S} \right) + 5\Omega^4 \Lambda = 0 , \tag{50}$$

and the $\delta w_A$-equation

$$\left( \Omega W^{AB} \right)_{:B} = 0 . \tag{51}$$

It is worth noting that contracting (49) one obtains equation (50), so that actually, the latter does not fix the 5D Dirac gauge function, and one is free to choose an arbitrary function $\Omega(x^A)$. (There is however, a "natural" restriction, $\Omega > 0$ on the gauge function.) Thus, one is left with only two field equations, (49) and (51).

One can consider in the bulk the Einstein gauge $\Omega = 1$, then from (49) and (51) follow the very simple equations

$$R_{AB} - \frac{1}{2} g_{AB} \hat{R} = 2\left( W_{AS} W_B{}^{\cdot S} - \frac{1}{4} g_{AB} W_{LS} W^{LS} \right) - \frac{1}{2} g_{AB} \Lambda , \tag{52}$$

$$\left(W^{AB}\right)_{;B} = 0 \ . \tag{53}$$

## 6. THE EQUATIONS ON THE 4D BRANE

In order to obtain equations of gravitation in the 4D brane we can make use of (49) and of the Gauss-Codazzi equations. The latter may be found in textbooks on differential geometry as well in works of Seahra and Wesson [47, 48]. The 5D Riemannian curvature tensor $R_{ABCD}$ is related to the 4D one $R_{\alpha\beta\gamma\delta}$ by the Gauss equation

$$R_{ABCD}\, e^A_\alpha e^B_\beta e^C_\gamma e^D_\delta = R_{\alpha\beta\gamma\delta} + 2\varepsilon\, C_{\alpha[\delta}C_{\gamma]\beta} \ , \tag{54}$$

where $C_{\alpha\beta}$ is the extrinsic curvature of the 4D brane $\Sigma_l$ in the 5D bulk (cf. (30)). There is also the Codazzi equation

$$R_{MABC}\, n^M e^A_\alpha e^B_\beta e^C_\gamma = 2C_{\alpha[\beta;\gamma]} \ . \tag{55}$$

Following [47, 48] we denote

$$E_{\alpha\beta} \equiv R_{MANB}\, n^M n^N e^A_\alpha e^B_\beta \ , \tag{56}$$

and introduce the contracted quantity (cf. (A-5))

$$E \equiv h^{\lambda\sigma} E_{\lambda\sigma} = -R_{MN}\, n^M n^N \ . \tag{57}$$

Then from (54) we obtain

$$R_{\alpha\beta} = e^A_\alpha e^B_\beta R_{AB} + \varepsilon\left[E_{\alpha\beta} - 2h^{\lambda\sigma} C_{\lambda[\sigma} C_{\beta]\alpha}\right] \ , \tag{58}$$

$$R = \hat{R} + 2\varepsilon\left[E - h^{\lambda\sigma} h^{\mu\nu} C_{\mu[\nu} C_{\lambda]\sigma}\right] \ , \tag{59}$$

with $R \equiv h^{\lambda\sigma} R_{\lambda\sigma}$ being the 4D curvature scalar, and $\hat{R} \equiv g^{MN} R_{MN}$ - the 5D one.

From (49), (50) one obtains the 5D Ricci tensor

$$R_{AB} = \frac{2}{\Omega^2}\left(W_{AS}W_B{}^{\cdot S} - \frac{1}{6}g_{AB}W_{NS}W^{NS}\right) + \frac{6}{\Omega^2}\left(\Omega_A\Omega_B - \frac{1}{3}g_{AB}\Omega_S\Omega^S\right) -$$
$$- \frac{3}{\Omega}\left(\Omega_{A:B} + \frac{1}{3}g_{AB}\Omega^S_{:S}\right) + \frac{g_{AB}}{3}\Omega^2\Lambda \ . \tag{60}$$

In addition to the notations given in (56), (57), we introduce the following:

$$B_{\alpha\beta} \equiv W_{AS}W_{BL}e^A_\alpha e^B_\beta n^S n^L, \text{ and } B = h^{\lambda\sigma}B_{\lambda\sigma} \equiv W_{AS}W_{BL}g^{AB}n^S n^L \ . \text{ (cf. (A-5))} \tag{61}$$

We also recall that $\Omega_{,A} \equiv \Omega_A$; $g^{AB}\Omega_{,B} \equiv \Omega^A$. Finally, substituting expression (60) and its contraction, $\hat{R} \equiv g^{AB}R_{AB}$, into equations (58, 59) and making use of (A-6) – (A-12) we obtain the 4D Einstein equation $G_{\alpha\beta} \equiv R_{\alpha\beta} - \frac{1}{2}h_{\alpha\beta}R = -8\pi T_{\alpha\beta}{}_{\text{total}}$ ,

$$G_{\alpha\beta} = -\frac{8\pi}{\Omega^2}M_{\alpha\beta} - \frac{2\varepsilon}{\Omega^2}\left(\frac{1}{2}h_{\alpha\beta}B - B_{\alpha\beta}\right) +$$
$$+ \frac{6}{\Omega^2}\Omega_\alpha\Omega_\beta - \frac{3}{\Omega}\left(\Omega_{\alpha;\beta} - h_{\alpha\beta}\Omega^\sigma_{;\sigma}\right) + \frac{3\varepsilon}{\Omega}\left(\Omega_S n^S\right)\left(h_{\alpha\beta}C - C_{\alpha\beta}\right) + \tag{62}$$
$$\varepsilon\left[E_{\alpha\beta} - h_{\alpha\beta}E + h^{\mu\nu}C_{\mu[\nu}C_{\lambda]\sigma}\left(h_{\alpha\beta}h^{\lambda\sigma} - 2\delta^\sigma_\alpha\delta^\lambda_\beta\right)\right] - \frac{1}{2}h_{\alpha\beta}\Omega^2\Lambda \ ,$$

with $M_{\alpha\beta} = \frac{1}{4\pi}\left(\frac{1}{4}h_{\alpha\beta}W_{\lambda\sigma}W^{\lambda\sigma} - W_{\alpha\lambda}W_\beta{}^{\cdot\lambda}\right)$ being the energy-momentum density tensor of the 4D electromagnetic field, and with $C \equiv h^{\lambda\sigma}C_{\lambda\sigma}$ .

On the right-hand-side of the gravitational equation (62) one discovers some interesting terms. The first line in addition to $M_{\alpha\beta}$ includes an induced by the bulk term that involves $W_{AL}W_{BS}$ of the bulk. This term may be interpreted as a relict of the 5D Weylian energy. The second line contains derivatives of the Dirac gauge function, where the third term of this line is induced by the bulk. In the third line appears the induced energy-momentum density tensor of the ordinary STM theory, as well a cosmological term.

Let us turn to the 5D Maxwell equation (51), $(\Omega W^{AB})_{;B} = 0$. From this we obtain the following 4D Maxwell equation (For details cf. (A-13) – (A-17) in the Appendix.)

$$W^{\alpha\beta}_{;\beta} = -\frac{\Omega_\beta}{\Omega} W^{\alpha\beta} + \varepsilon n_S \left[ W^{AS}\left(e^\beta_A h^{\alpha\lambda} - e^\alpha_A h^{\beta\lambda}\right) C_{\beta\lambda} + n^C e^\alpha_A \left(W^{AS}_{:C} + W^{AS}\frac{\Omega_C}{\Omega}\right) \right] . \quad (63)$$

From (63) one has the 4D electromagnetic current vector $J^\alpha = \frac{1}{4\pi} W^{\alpha\beta}_{;\beta}$, which includes a gauge depending term and an induced term. By symmetries one has

$$\left(\frac{\Omega_\beta}{\Omega} W^{\alpha\beta}\right)_{;\alpha} = 0 , \quad (64)$$

so that from the current conservation law we have the condition

$$\left[ \varepsilon n_S W^{AS}\left(e^\beta_A h^{\alpha\lambda} - e^\alpha_A h^{\beta\lambda}\right) C_{\beta\lambda} + \varepsilon n_B n^C e^\alpha_A \left(W^{AB}_{:C} + W^{AB}\frac{\Omega_C}{\Omega}\right) \right]_{;\alpha} = 0 . \quad (65)$$

In the Einstein gauge, $\Omega = 1$, one has from (62) and (63) the simple equations

$$G_{\alpha\beta} = -8\pi M_{\alpha\beta} - 2\varepsilon\left(\frac{1}{2} h_{\alpha\beta} B - B_{\alpha\beta}\right) +$$
$$\varepsilon\left[E_{\alpha\beta} - h_{\alpha\beta} E + h^{\mu\nu} C_{\mu[\nu} C_{\lambda]\sigma}\left(h_{\alpha\beta} h^{\lambda\sigma} - 2\delta^\sigma_\alpha \delta^\lambda_\beta\right)\right] - \frac{1}{2} h_{\alpha\beta}\Omega^2 \Lambda , \quad (66)$$

and (cf. (A-17))

$$W^{\alpha\beta}_{;\beta} = \varepsilon n_S \left[W^{AS}\left(e^\beta_A h^{\alpha\lambda} - e^\alpha_A h^{\beta\lambda}\right) C_{\beta\lambda} + n^C e^\alpha_A W^{AS}_{:C}\right] . \quad (67)$$

In absent of the fifth dimension ($\varepsilon = 0$) and in the Einstein gauge $\Omega = 1$, one is left with

$$G_{\alpha\beta} = -8\pi M_{\alpha\beta} - h_{\alpha\beta}\tilde{\Lambda}; \text{ where } \tilde{\Lambda} \equiv \frac{1}{2}\Lambda , \quad (68)$$

and with

$$W^{\alpha\beta}_{;\beta} = 0 . \quad (69)$$

For cosmology the case $W_{AB} = 0$, but with an arbitrary positive Dirac gauge function may be of interest. Here, from (62) one has

$$G_{\alpha\beta} = \frac{6}{\Omega^2}\Omega_\alpha\Omega_\beta - \frac{3}{\Omega}\left(\Omega_{\alpha;\beta} - h_{\alpha\beta}\Omega^\sigma_{;\sigma}\right) - \frac{1}{2}h_{\alpha\beta}\Omega^2\Lambda + \varepsilon\frac{3}{\Omega}\left(\Omega^S n_S\right)\left(h_{\alpha\beta}C - C_{\alpha\beta}\right) +$$
$$+ \varepsilon\left[\left(E_{\alpha\beta} - h_{\alpha\beta}E\right) - C\left(C_{\alpha\beta} - \frac{1}{2}h_{\alpha\beta}C\right) + h^{\lambda\sigma}\left(C_{\alpha\lambda}C_{\beta\sigma} - \frac{1}{2}h_{\alpha\beta}h^{\mu\nu}C_{\mu\lambda}C_{\nu\sigma}\right)\right] \quad . \tag{70}$$

If we for a moment assume that $\varepsilon = 0$, we obtain from (70) the equation

$$G_{\mu\nu} = \frac{6}{\Omega^2}\Omega_\mu\Omega_\nu - \frac{3}{\Omega}\left(\Omega_{\mu;\nu} - g_{\mu\nu}\Omega^\sigma_{;\sigma}\right) - \frac{1}{2}g_{\mu\nu}\Omega^2\Lambda \quad . \tag{71}$$

This may be regarded as describing a universe filled with the Dirac gauge function and a cosmological term.

## 7. DISCUSSION

In the present paper, we have discussed Wesson's Induced Matter Theory (cf. [1, 47, 48]) and proposed a Weyl-Dirac version of this theory.

Suppose one carries out an infinitesimal parallel displacement of a vector $V_A$ in the 5D Riemannian bulk $\{M\}$. Then, the change in the length $\left(V^S V_S\right)^{\frac{1}{2}}$ obviously vanishes. However, considering the induced parallel displacement of the 4D counterpart $V_\alpha = e^A_\alpha V_A$ in the 4D hypersurface (brane), one discovers that its length $\left(V^\sigma V_\sigma\right)^{\frac{1}{2}}$ changes (cf. (32)), so that the brane is no longer a Riemannian space. The mentioned change is induced by the bulk and involves the external curvature of the 4D brane.

This non-integrability of the induced 4D geometry justifies revising the geometry of the bulk. It is shown in the present paper that the pure geometric 5D bulk is a Weylian space rather than a Riemannian one. Therefore, we have built up a Weyl-Dirac formalism on the bulk. The 5D manifold $\{M\}$ is mapped by coordinates $\{x^N\}$ and in every point exist the Weylian connection vector $w^A$, the Dirac gauge function $\Omega$, and the symmetric metric tensor $g_{AB}$. The three fields $g_{AB}$, $w^A$ and $\Omega$ are integral parts of the geometric framework, and there are no additional fields or particles in $\{M\}$. In the bulk, two field equations, for $g_{AB}$ and $w^A$, are derived from a geometrically based action, whereas the Dirac gauge function $\Omega$ may be chosen arbitrarily. Making use of the Gauss-Codazzi equations we obtained the equations in the 4D hypersurface (brane), one for the gravitational field (cf. (62)) and another describing the $w_\mu$-field, the latter being a generalization of the Maxwell equation (cf. (63)). The sources in both equations consist of two parts: the first is located in the brane and it depends on the gauge function and on the 4D Maxwell field tensor, the second is induced by the bulk.

Some interesting cases may be noted:

1. If the external curvature of the brane vanishes ($C_{\alpha\beta} \equiv e_\alpha^A e_\beta^B (n_B)_{;A} = 0$) one has from equation (62)

$$G_{\alpha\beta} = -\frac{8\pi}{\Omega^2} M_{\alpha\beta} - \frac{2\varepsilon}{\Omega^2}\left(\frac{1}{2}h_{\alpha\beta}B - B_{\alpha\beta}\right) + \frac{6}{\Omega^2}\Omega_\alpha\Omega_\beta \qquad (62a)$$
$$-\frac{3}{\Omega}\left(\Omega_{\alpha;\beta} - h_{\alpha\beta}\Omega^\sigma_{;\sigma}\right) + \varepsilon\left[E_{\alpha\beta} - h_{\alpha\beta}E\right] - \frac{1}{2}h_{\alpha\beta}\Omega^2\Lambda ,$$

and (63) may be written as

$$W^{\alpha\beta}_{;\beta} = -\frac{\Omega_\beta}{\Omega}W^{\alpha\beta} + \varepsilon n_S n^C e_A^\alpha \left(W^{AS}_{:C} + \frac{\Omega_C}{\Omega}W^{AS}\right) . \qquad (63a)$$

Further, we can turn to the Einstein gauge ($\Omega = 1$), then

$$G_{\alpha\beta} = -\frac{8\pi}{\Omega^2} M_{\alpha\beta} - \frac{2\varepsilon}{\Omega^2}\left(\frac{1}{2}h_{\alpha\beta}B - B_{\alpha\beta}\right) + \varepsilon\left[E_{\alpha\beta} - h_{\alpha\beta}E\right] - \frac{1}{2}h_{\alpha\beta}\Omega^2\Lambda . \qquad (62b)$$

If in addition we impose the four conditions: $n_S n^C e_A^\alpha W^{AS}_{:C} = 0$, we are left with

$$W^{\alpha\beta}_{;\beta} = 0. \qquad (63b)$$

Equation (63b) describes electromagnetic radiation in a free of sources 4D space-time. In cosmology (62b), (63b) present a universe filled with radiation and induced matter.

2. If the 5-dimensional Weylian vector is a gradient vector ($w_A = \dfrac{\partial f(x^N)}{\partial x^A}$) one has in the bulk an Integrable Weyl-Dirac geometry. Then on the 4D brane there is no electromagnetism, whereas the gravitational field is given by equation (70). In the latter one can recognize three sources, the first formed from the Dirac gauge function, the second including the cosmological term, the third being induced by the bulk. This case, which presents a universe filled with matter and a cosmic scalar field, may be of great interest for constructing cosmological models. If in addition one turns to the Einstein gauge ($\Omega = 1$), he gets the equation of the ordinary STM theory (cf. [47, 48]).

The proposed Weyl-Dirac version of Wesson's IMT is justified from the geometrically cognitive point of view (cf. sec. 4, 5). This modified IMT opens new possibilities of building up geometrically based unified theories of gravitation and electromagnetism. It also may be very useful for obtaining theoretical scenarios that involve dark matter and quintessence, as well for singularity-free cosmological models. These problems will be considered in subsequent works.

Finally, it must be pointed out that in the proposed framework one is faced by the non-integrability of length, which causes difficulties in fixing measuring standards. There are, however, ways of overcoming this obstacle, so that one can define a unique measuring standard. Suitable procedures are developed in a forthcoming paper.

# APPENDIX A

**1)** Below we give variations of various terms in the action (48). The first term:

$$\delta\left(\Omega W_{AB} W^{AB} \sqrt{-g}\right) = \left(W_{AB} W^{AB} \sqrt{-g}\right)\delta\Omega - 4\left(\Omega W^{AB}\right)_{:B} \sqrt{-g}\, \delta w_A +$$
$$+ 2\Omega\left(W_{AM} W_B^{\ M} - \frac{1}{4} g_{AB} W_{LS} W^{LS}\right)\sqrt{-g}\, \delta g^{AB} \, . \qquad (A\text{-}1)$$

The variation of the second term in (48) yields

$$\delta\left(\Omega^3 \hat{R}\sqrt{-g}\right) = 3\Omega^2 \hat{R}\sqrt{-g}\, \delta\Omega +$$
$$+ \left[\Omega^3\left(R_{AB} - \frac{1}{2} g_{AB} \hat{R}\right) + 3\left(\Omega^2 \Omega_A\right)_{:B} - 3 g_{AB}\left(\Omega^2 \Omega^S\right)_{:S}\right]\sqrt{-g}\, \delta g^{AB} \, . \qquad (A\text{-}2)$$

For the third term we obtain

$$\delta\left(\Omega \Omega_S \Omega^S \sqrt{-g}\right) = \left[\Omega_S \Omega^S - 2\left(\Omega \Omega^S\right)_{:S}\right]\sqrt{-g}\, \delta\Omega +$$
$$+ \Omega\left(\Omega_A \Omega_B - \frac{1}{2} g_{AB} \Omega_S \Omega^S\right)\sqrt{-g}\, \delta g^{AB} \, . \qquad (A\text{-}3)$$

Finally, varying the cosmological term we obtain

$$\delta\left(\Omega^5 \Lambda \sqrt{-g}\right) = 5\Omega^4 \Lambda \sqrt{-g}\, \delta\Omega - \frac{1}{2} g_{AB} \Omega^5 \Lambda \sqrt{-g}\, \delta g^{AB} \, . \qquad (A\text{-}4)$$

From (56) one has:

$$E_{\lambda\sigma} h^{\lambda\sigma} = R_{MANB} n^M n^N \left(g^{AB} - \varepsilon n^A n^B\right) = -R_{AMNB}\, g^{AB} n^M n^N \equiv -R_{MN} n^M n^N \, . \quad (A\text{-}5)$$

**2)** Below are given some simple relations, which were used in the process of deriving the 4D Einstein equation (62).

For any tensor having two indices one has $T^{\alpha\beta} = T^{AB} e_A^\alpha e_B^\beta$ multiplying this by $e_\alpha^M e_\beta^N$ one obtains the following decomposition of $T^{AB}$:

$$T^{AB} = T^{\alpha\beta} e_\alpha^A e_\beta^B + \varepsilon\left(T^{AS} n^B + T^{SB} n^A\right) n_S - T^{SL} n_S n_L n^A n^B \, , \qquad (A\text{-}6)$$

and for an antisymmetric tensor this gives

$$W^{AB} = W^{\alpha\beta} e_\alpha^A e_\beta^B + \varepsilon\left(W^{AS} n^B + W^{SB} n^A\right) n_S \, , \qquad (A\text{-}7)$$

$$W_A{}^{\cdot B} = W_\alpha{}^{\cdot \beta} e_A^\alpha e_\beta^B + \varepsilon \left(W_S{}^{\cdot B} n^S n_A + W_A{}^{\cdot S} n^B n_S\right). \tag{A-8}$$

From (A-7), (A-8) one obtains

$$e_\alpha^A e_\beta^B W_{AS} W_B{}^{\cdot S} = W_{\alpha\sigma} W_\beta{}^{\cdot \sigma} + \varepsilon\, W_{AS} W_B{}^{\cdot L} n^S n_L e_\alpha^A e_\beta^B, \tag{A-9}$$

$$W_{AB} W^{AB} = W_{\alpha\beta} W^{\alpha\beta} + 2\varepsilon\, W_{AS} W^{AL} n^S n_L. \tag{A-10}$$

Two more relations are

$$\Omega^S \Omega_S = \Omega^\sigma \Omega_\sigma + \varepsilon \left(\Omega^L n_L\right)^2, \tag{A-11}$$

$$\Omega^S{}_{:S} = \Omega^\sigma{}_{;\sigma} + \varepsilon \Omega^S n_S\, h^{\sigma\lambda} C_{\sigma\lambda}. \tag{A-12}$$

**3)** In order to get from (51) its 4D counterpart we prove first the following simple relation:

$$W_{AB} e_\alpha^A e_\beta^B = W_{\alpha\beta}. \tag{A-13}$$

Generally

$$e_\alpha^A e_\beta^B \left(w_{A:B} - w_{B:A}\right) = w_{\alpha;\beta} - w_{\beta;\alpha} + \varepsilon \left(w_S n^S\right) e_\alpha^A e_\beta^B \left(n_{A:B} - n_{B:A}\right) \tag{A-14}$$

The last term in (A-14) may be rewritten as $\varepsilon\left(w_S n^S\right)\left(C_{\alpha\beta} - C_{\beta\alpha}\right)$ and it vanishes as $C_{\alpha\beta} = C_{\beta\alpha}$. Thus, (A-13) is right.

Now, let us consider the Riemannian (cf. (41)) derivative of a 5D vector $V^A$.

By a straightforward calculation one obtains

$$e_A^\alpha e_\gamma^C \left(V^A\right)_{:C} = V^\alpha{}_{;\gamma} + \varepsilon \left(V^S n_S\right) h^{\alpha\lambda} C_{\lambda\gamma}. \tag{A-15}$$

The enlargement of (A-15) for $W^{AB}$ is

$$e_A^\alpha e_B^\beta e_\gamma^C \left(W^{AB}\right)_{:C} = W^{\alpha\beta}{}_{;\gamma} + \varepsilon n_S \left[W^{SB} e_B^\beta h^{\alpha\lambda} + W^{AS} e_A^\alpha h^{\beta\lambda}\right] C_{\lambda\gamma}. \tag{A-16}$$

It is interesting that the structure of (A-16) resembles that of the Gauss equation (54).
Finally, multiplying (A-16) by $\delta_\beta^\gamma$ we obtain the 4D Maxwell equation in the Einstein gauge (67)

$$W^{\alpha\beta}_{;\beta} = \varepsilon n_S W^{AS}\left(e_A^\beta h^{\alpha\lambda} - e_A^\alpha h^{\beta\lambda}\right)C_{\beta\lambda} + \varepsilon n_B n^C e_A^\alpha W^{AB}_{:C} \qquad (A-17)$$

From this one easily obtain the Maxwell equation (63) with arbitrary gauges.

## APPENDIX B

In this section we consider some details concerned with Weylian differentiation. There is a simple relation between the 5D and 4D Weylian connections

$$e_\alpha^A e_\beta^B e_C^\gamma (g_{AB} w^C - \delta_A^C w_B - \delta_B^C w_A) = h_{\alpha\beta} w^\gamma - \delta_\alpha^\gamma w_\beta - \delta_\beta^\gamma w_\alpha . \qquad (B-1)$$

For a given 5D vector $V_A$ the 5D Weylian derivative (cf. (34), (40), and (41)) is

$$\nabla_B V_A = V_{A:B} - V_S\left(g_{AB} w^S - \delta_A^S w_B - \delta_B^S w_A\right), \qquad (B-2)$$

and for the 4D counterpart $V_\alpha \equiv e_\alpha^A V_A$ one has the 4D Weylian derivative

$$\nabla_\beta V_\alpha = V_{\alpha;\beta} - V_\sigma\left(h_{\alpha\beta} w^\sigma - \delta_\alpha^\sigma w_\beta - \delta_\beta^\sigma w_\alpha\right). \qquad (B-3)$$

By a straightforward procedure, one obtains the following relation:

$$e_\alpha^A e_\beta^B \nabla_B V_A = \nabla_\beta V_\alpha + \varepsilon\left(V_S n^S\right)\left[C_{\alpha\beta} - h_{\alpha\beta}\left(w_R n^R\right)\right] . \qquad (B-4)$$

Thus, in addition to the 4D Weylian derivative appears a term induced by the bulk. One can however consider the following somewhat artificial and questionable approach. Let us take the projection tensor $h_{AB} = g_{AB} - \varepsilon n_A n_B$; $h_A^C = \gamma_A^C - \varepsilon n_A n^C$ (cf. [47, 48]). Instead of $V_A$ we consider its 5D projection $h_A^C V_C$. With this modified vector we obtain

$$e_\alpha^A e_\beta^B \left(h_A^C V_C\right)_{:A} = V_{\alpha;\beta} \:, \tag{B-5}$$

and

$$e_\alpha^A e_\beta^B \nabla_B \left(h_A^C V_C\right) = \nabla_\beta V_\alpha \:. \tag{B-6}$$

Here the effect of the bulk is eliminated. This is not astonishing, as we have taken the projection on the hypersurface instead of the vector itself.